\documentclass[prb,twocolumn,superscriptaddress,showpacs]{revtex4}

\usepackage[dvipdfm]{graphicx}
\usepackage{epsbox}
\usepackage{graphicx}
\usepackage{threeparttable}
\usepackage{txfonts}
\usepackage{bm}
\usepackage{color}
\usepackage{hyperref}
\usepackage{enumerate}

\begin{document}
\title{Electrostatics-based finite-size correction for first-principles point defect calculations}

\author{Yu Kumagai}
\email[]{yuuukuma@gmail.com}
\affiliation{Materials Research Center for Element Strategy, Tokyo Institute of Technology, Yokohama 226-8503, Japan}

\author{Fumiyasu Oba}
\affiliation{Materials Research Center for Element Strategy, Tokyo Institute of Technology, Yokohama 226-8503, Japan}
\affiliation{Department of Materials Science and Endineering, Kyoto University, Kyoto 606-8501, Japan}

\date{\today}

\definecolor{blue}{rgb}{0.0,0.0,1}
\hypersetup{colorlinks,breaklinks,linkcolor=blue,urlcolor=blue,anchorcolor=blue,citecolor=blue}

\begin{abstract}
Finite-size corrections for charged defect supercell calculations typically consist of image-charge and potential alignment corrections.
A wide variety of schemes for both corrections have been proposed for decades.
Regarding the image-charge correction, Freysoldt, Neugebauer, and Van de Walle (FNV) recently proposed a novel method that enables us to accurately estimate the correction energy 
{\it a posteriori} through alignment of the defect-induced potential to the model charge potential [C. Freysoldt, J. Neugebauer, and C. G. Van de Walle, Phys. Rev. Lett. 102, 016402 (2009).]
This method, however, still has two issues in practice.
Firstly, it uses planar-averaged electrostatic potential for determining the potential offset, which cannot be readily applied to relaxed atomic structure.
Secondly, the long-range Coulomb interaction is assumed to be screened by a macroscopic dielectric constant.
This is valid only for cubic systems and can bring forth huge errors for defects in anisotropic materials, particularly with layered and low-dimensional structures.
In the present study, we use the atomic site electrostatic potential as a potential marker instead of the planar-averaged potential, 
and extend the FNV scheme by adopting the point charge model in an anisotropic medium for estimating long-range interactions.
We also revisit the conventional potential alignment correction and show that it is fully included in the image-charge correction and therefore unnecessary.
In addition, we show that the potential alignment corresponds to a part of first-order and full of third-order image-charge correction;
thus the third-order image-charge contribution is absent after the potential alignment.
Finally, a systematic assessment of the accuracy of the extended FNV correction scheme is performed for a wide range of material classes: 
$\beta$-Li$_2$TiO$_3$, ZnO, MgO, corundum Al$_2$O$_3$, monoclinic HfO$_2$, cubic and hexagonal BN, Si, GaAs, and diamond.
The defect formation energies with -6 to +3 charges calculated using around 100-atom supercells are successfully corrected even after atomic relaxation 
within a few tenths of eV compared to those in the dilute limit.
\end{abstract}


\maketitle
\section{Introduction: First-principles calculations of point defects and their correction schemes}
Point defects and impurities are ubiquitous in semiconductors or insulators and strongly dominate a wide variety of materials properties 
such as optical, mechanical, electrical, and transport properties, 
having a decisive impact on their performance in applications, e.g. photovoltaics, photocatalysts, ionic conductors, transistors, and light emitting diodes.
Therefore, knowledge and precise control of defects are inherently keys to the smart design of materials with superior performance. 
Despite the importance, it is difficult to directly and fully study point defects by experiments,
and first-principles calculations have emerged as an invaluable tool for modeling and understanding the point defects.~\cite{VandeWalle:2004,Nieminen:2009bt,Lambrecht:2010jt} 
In particular, a rapid progress on the computational speed and electronic structure calculation methods as represented by hybrid functionals, quantum Monte Carlo, and the GW approximation 
allows us to predict the defect properties semi-quantitatively or even quantitatively in recent years.~\cite{RevModPhys.73.33,PhysRevB.63.075205,PhysRevB.74.121102,PhysRevB.77.245202,PhysRevLett.101.046405,PhysRevLett.103.176403,PhysRevLett.102.026402,PhysRevB.81.113201,Chen:2012tj,Gruneis.2014}
These calculations support and complement experimental findings.

The first-principles point defect calculations commonly rely on the supercell approach under periodic boundary conditions.
However, the cell sizes are not usually sufficiently large for describing the low concentration of defects in realistic materials such as 10$^{14}$  -- 10$^{18}$ cm$^{-3}$.
Calculations using common approximations to density functional theory (DFT), viz. local density approximation (LDA) or generalized gradient approximation (GGA), 
can treat a few thousand atoms at most, and hybrid functionals such as the Heyd-Scuseria-Ernzerhof functional (HSE06)~\cite{JChemPhys.118.8207,JChemPhys.125.224106} up to a few hundred atoms, 
which corresponds to 10$^{20}$ -- 10$^{21}$ cm$^{-3}$.
It is notorious that the formation energy of the charged defect calculated with such smaller supercells could include huge convergence errors up to several eV. 
In case that the defect charge is encased in the supercell, 
the main source of the error comes from the spurious long-range Coulomb interactions between the defect charge, its periodic images and background charge,~\cite{Lany:2008gk,Komsa:2012ji}
which is requisite for avoiding the divergence of the electrostatic energy.
Consequently, the formation energy slowly converges with the supercell size.
A correction for image-charge interactions is therefore inevitable for evaluating the defect formation energy in the isolated limit 
unless the dielectric constant is large enough to screen the spurious interactions.
In addition, since the average electrostatic potential in the entire supercell is conventionally set to zero within the momentum-space formalism,
the eigenvalues are defined only up to an undetermined constant.~\cite{Kleinman:1981fn}
Whereas the total energy of a charge-neutral system is well defined, the charged system depends on the undetermined shift of the valence band maximum (VBM).
Therefore, it has been believed that one needs to align the VBM in the calculations of charged defects to that of the pristine host 
for restoring physically meaningful formation energies.~\cite{VandeWalle:2004,Lany:2009vf}
This is a so-called potential alignment correction.

The formation energy of defect $D$ in charge state $q$ is estimated as~\cite{PhysRevLett.67.2339,Komsa:2012ji}
\begin{eqnarray}
E_f[D^q] &=& \Bigl\{E[D^q] + E_{\rm corr}[D^q]\Bigr\} - E_P -  \sum n_i\mu_i \nonumber \\
 &+& q \Bigl\{ (\epsilon_{\rm VBM}+\Delta v) + \Delta\epsilon_F \Bigr\}.
\label{defect_formation_energy}
\end{eqnarray}
Here $E[D^q]$ and $E_P$ are the total energies of the supercell with the defect $D$ in charge state $q$ and the perfect supercell without defect, respectively. 
$n_i$ is the number of removed ($n_i$ $<$ 0) or added ($n_i$ $>$ 0) $i$-type atom and $\mu_i$ refers to the chemical potential.
$\epsilon_{\rm VBM}$ is the energy level of the VBM, and $\Delta\epsilon_F$ is the Fermi level referenced to $\epsilon_{\rm VBM}$.
$E_{\rm corr}[D^q]$ and $\Delta v$, corresponding to the image-charge correction and potential alignment correction, respectively, are for charged defects.
Then $\epsilon_{\rm VBM}+\Delta v+\Delta\epsilon_F$ ($=\epsilon_F$) represents the Fermi level.

A number of image-charge correction schemes have been proposed since a few decades ago.
~\cite{Leslie:1985bn,Makov:1995vk,PhysRevLett.84.1942,PhysRevB.73.035215,Lany:2008gk,Freysoldt:2009ih,Freysoldt:2011hf,Taylor:2011ue}
The simplest correction is the point charge (PC) correction, which is a leading term for correcting spurious electrostatic interactions.
Unfortunately, in some cases the higher-order terms are not negligible,
and then the defect formation energy has to be extrapolated to the infinite interdefect distance limit with a set of supercell calculations.
This is, however, prohibited when computational costs severely limit the size of supercells as seen in hybrid functional calculations.

Recently, Freysoldt, Neugebauer, and Van de Walle (FNV) proposed a remarkable scheme, 
which allows us to correct the defect formation energies {\it a posteriori}.~\cite{Freysoldt:2009ih,Freysoldt:2011hf}
A great advantage of this scheme is to estimate a correction energy from two supercell calculations with and without a defect and require no additional first-principles calculations.
Therefore it is useful especially when computationally expensive methods are employed.
This correction scheme, however, still has two practical issues.

Firstly, it uses planar-averaged electrostatic potential for aligning the defect-induced potential, which is obtained by subtracting bulk supercell potential from defective supercell potential, 
to the model charge potential.
This works well when the atomic positions are fixed.
However, realistic defect calculations require relaxing the atomic positions, 
and the defect-induced potential becomes scraggly because of the atomic displacements.
Consequently, the potential offset between defect-induced potential and model charge potential 
has to be determined, e.g. by convoluting the defect-induced potential with a suitable Gaussian function.~\cite{Komsa:2012ji}
Secondly, in the original paper, the long-range Coulomb interaction is assumed to be screened by a macroscopic dielectric constant.
This is valid only for cubic systems, and the dielectric constant should be replaced by a dielectric tensor for other systems.
In order to resolve these two practical issues and make FNV scheme applicable to broad classes of materials, 
we use atomic site electrostatic potential for evaluating the defect-induced potential and an anisotropic PC model for long-range Coulomb interactions.
The extended FNV scheme is applied to the layered compounds, hexagonal BN (h-BN) and $\beta$-Li$_2$TiO$_3$, as well as three-dimensional systems.
Details are discussed in Sec.~\ref{image}.

There also exist a wide variety of fashions for the potential alignment.
In the most-used way, the potential is aligned so that the electrostatic potential at the outermost atomic sites in the supercell with a charged defect 
becomes the same as that of the bulk.~\cite{VandeWalle:2004,PhysRevB.72.035211,PhysRevB.80.085120,Choi:2013uo}
Instead, Lany and Zunger adopted the reference by averaging potential differences from the perfect cell at all atomic sites except the immediate neighbors of defects.~\cite{Lany:2009vf}
Taylor and Bruneval, however, demonstrated that the Madelung potential, which is taken into account by the first-order image-charge correction, brings a potential shift
and one cannot perform the image-charge correction and potential alignment independently.~\cite{Taylor:2011ue}
In order to remove the long-range Coulomb interactions, Komsa {\it et al}. proposed a way to align the potential at the outermost area of the {\it neutral} defect
to that of the pristine bulk.~\cite{Komsa:2012ji}
Taylor and Bruneval also proposed to align the potential averaged over the entire supercell including exchange-correlation (XC) potential to the bulk potential.~\cite{Taylor:2011ue}
In Sec.~\ref{alignment_revisited}, we revisit the controversial potential alignment, 
and conclude that the potential alignment is unnecessary ($\Delta v = 0$) as long as the image-charge correction is properly adopted.

To our best knowledge, the cell size dependence of the FNV correction for relaxed defects has been reported only by Komsa {\it et al.} with $V_{\rm O}^{+1}$ in MgO.~\cite{Komsa:2012ji}
To assess the performance of the correction scheme is essential for practical applications.
In Sec.~\ref{test}, we apply the extended FNV scheme introduced in this study to a wide variety of material classes:
ZnO, MgO, corundum Al$_2$O$_3$, monoclinic HfO$_2$ (m-HfO$_2$), cubic BN (c-BN), Si, GaAs, and diamond in addition to $\beta$-Li$_2$TiO$_3$ and h-BN with layered structures
and estimate its accuracy for relaxed defects.
In addition, we discuss the remaining error sources.

\section{Details of first-principles calculations}\label{calc_condition}
We here summarize the details of the first-principles calculations used in this study.
Our calculations were performed using the projector augmented-wave (PAW) method\cite{PhysRevB.50.17953} as implemented in {\sc vasp}.\cite{PhysRevB.47.558,PhysRevB.54.11169}
We adopted Perdew-Burke-Ernzerhof GGA (PBE-GGA)~\cite{PhysRevLett.77.3865} except for GaAs and diamond:
GaAs was calculated with the LDA~\cite{PhysRevB.23.5048} because the band gap is significantly underestimated 
with the PBE-GGA at the equilibrium lattice constant (0.16 eV with the GGA vs. 0.51 with the LDA), and
diamond was calculated with the HSE06 hybrid functional for demonstrating the correction of HSE06 defect formation energy.
A Hubbard $U$ correction was applied to Ce in c-BN ($U-J$ = 4.5 eV for $f$ orbitals).~\cite{PhysRevB.57.1505,PhysRevLett.110.065504}

In this study, 
Li 2$s$,
B 2$s$ and 2$p$,
C 2$s$ and 2$p$,
N 2$s$ and 2$p$,
O 2$s$ and 2$p$,
Mg 3$s$,
Al 3$s$ and 3$p$,
Si 3$s$ and 3$p$,
Ti 4$s$ and 3$d$,
Zn 4$s$ and 3$d$,
Ga 4$s$ and 4$p$,
As 4$s$ and 4$p$, and
Ce 4$f$, 5$d$, and 6$s$,
and Hf 6$s$ and 5$d$
were described as valence electrons.
The PAW data set with radial cutoffs of 1.08, 0.90, 0.70, 0.79, 0.80, 1.06, 1.01, 1.01, 1.48, 1.22, 1.38, 1.11, 1.36, and 1.59 $\AA$ was used 
for Li, B, C, N, O, Mg, Al, Si, Ti, Zn, Ga, As, Ce, and Hf, respectively.
The average atomic site potential was evaluated within spheres of radii 0.97, 0.77, 0.79, 0.71, 0.72, 1.07, 1.04, 0.99, 1.28, 1.06, 1.26, 0.95, and 1.25 $\AA$
for Li, B, C, N, O, Mg, Al, Si, Ti, Zn, Ga, As, and Hf.
Wave functions were expanded with plane waves up to energy cutoffs of 400 and 550 eV for the cases where lattice parameters were fixed and optimized, respectively.
Integrations in reciprocal space were performed with $\Gamma$-centered grids so that the total energies sufficiently converge.
In this study, atomic positions were relaxed, but the lattice parameters were fixed at the bulk optimized values for defect calculations unless otherwise noted.
Forces acting on the atoms and stresses were reduced to be less than 0.02 eV/$\AA$ and 0.05 GPa.
The dielectric tensors are indispensable for the correction of the defect formation energies.
Both ion-clamped dielectric tensors and ionic contributions to the dielectric tensors were calculated with density functional perturbation theory.~\cite{PhysRevB.33.7017,PhysRevB.73.045112}

The calculated lattice parameters and dielectric tensors are summarized in Table ~\ref{list_properties}.
The lattice constants estimated with the PBE-GGA are systematically overestimated, which is a typical tendency in the PBE-GGA.
The ion-clamped dielectric constants are overestimated compared to the experimental ones except for diamond that is treated using the HSE06.
This would be related to underestimation of the calculated band gaps with the LDA and PBE-GGA.
Note that only an ion-clamped dielectric tensor, 
and the sum of an ion-clamped dielectric tensor and an ionic contribution should be used for the correction of unrelaxed and relaxed systems, respectively.~\cite{Komsa:2012ji}

\begin{table}
 \caption{
   Calculated lattice parameters in units of $\AA$ and degrees, the ion-clamped ($\epsilon^{\rm ele}$) macroscopic dielectric tensors 
and ionic contributions to the dielectric tensors ($\epsilon^{\rm ion}$)
in ZnO (space group: $P6_3mc$), MgO ($Fm\overline{3}m$), Al$_2$O$_3$ ($R\overline{3}c$), HfO$_2$ ($P2_1/c$),
c-BN ($F\overline{4}3m$), h-BN ($P6_3/mmc$), $\beta$-Li$_2$TiO$_3$ ($C2/c$), Si ($Fd3m$), GaAs ($F\overline{4}3m$), and diamond ($Fd3m$).
Available experimental values are also shown.
The experimental $\epsilon^{\rm ele}$ are high-frequency dielectric constants $\epsilon^{\infty}$, 
and $\epsilon^{\rm ion}$ are estimated by subtracting $\epsilon^{\infty}$ from static dielectric constants. 
Note that ionic contributions of elemental substances (Si and diamond) are null because Born effective charges are zero.}
  \begin{center}
  \begin{threeparttable}
  \begin{tabular}{cccc}\\ 
  \hline
  \hline
     & Lattice param. & $\epsilon^{\rm ele}$ & $\epsilon^{\rm ion}$ \\
  \hline
ZnO&$\begin{array}{c}a=3.29\\b=5.31\end{array}$&$\begin{array}{c} \epsilon_{\perp}=5.20\\\epsilon_{\parallel}=5.22 \end{array}$&$\begin{array}{c}\epsilon_{\perp}=5.14\\\epsilon_{\parallel}=6.02 \end{array}$\\
exp.\footnote{References \onlinecite{Albertsson:st0267,Ashkenov:2003kq}} &$\begin{array}{c}a=3.250\\b=5.207\end{array}$&$\begin{array}{c} \epsilon_{\perp}=3.70\\\epsilon_{\parallel}=3.78 \end{array}$&$\begin{array}{c}\epsilon_{\perp}=4.07\\\epsilon_{\parallel}=5.13 \end{array}$\\
  \hline
MgO  & $a=4.25$&3.16&7.50\\
 exp.\footnote{References \onlinecite{Taylor1984,Komsa:2012ji}} &$a=4.211$&3.0&6.6\\
  \hline
Al$_2$O$_3$&$\begin{array}{c}a=4.81\\c=13.12\end{array}$&$\begin{array}{c} \epsilon_{\perp}=3.27\\\epsilon_{\parallel}=3.24 \end{array}$&$\begin{array}{c}\epsilon_{\perp}=6.74\\\epsilon_{\parallel}=9.11 \end{array}$\\
  exp.\footnote{References \onlinecite{ZKristallogr.117.235,PhysRevB.61.8187}}  &$\begin{array}{c}a=4.76\\c=12.99\end{array}$&$\begin{array}{c} \epsilon_{\perp}=3.1\\\epsilon_{\parallel}=3.1 \end{array}$&$\begin{array}{c}\epsilon_{\perp}=6.3\\\epsilon_{\parallel}=8.5 \end{array}$\\
  \hline
HfO$_2$ &$\begin{array}{c}a=5.14\\b=5.19\\c=5.32\\\beta = 100\end{array}$&$\begin{array}{c}\epsilon_{11}=4.79\\\epsilon_{22}=4.77\\\epsilon_{33}=4.52\\\epsilon_{13}=-0.13\\\end{array}$&
                                                                                $\begin{array}{c}\epsilon_{11}=15.17\\\epsilon_{22}=13.46\\\epsilon_{33}=10.74\\\epsilon_{13}=-1.09\\\end{array}$\\
exp.\footnote{Reference \onlinecite{JACE:JACEC-285}} &$\begin{array}{c}a=5.117\\b=5.175\\c=5.292\\\beta =99 \end{array}$ & NA & NA\\

  \hline
$\beta$-Li$_2$TiO$_3$ &$\begin{array}{c}a=5.09\\b=8.85\\c=9.82\\\beta = 100\end{array}$&$\begin{array}{c}\epsilon_{11}=5.45\\\epsilon_{22}=5.49\\\epsilon_{33}=3.74\\\epsilon_{13}=0.01\\\end{array}$&
                                                                                     $\begin{array}{c}\epsilon_{11}=36.95\\\epsilon_{22}=36.32\\\epsilon_{33}=12.07\\\epsilon_{13}=-1.06\\\end{array}$\\
exp.~\footnote{Reference \onlinecite{MaterResBull.44.168}} &$\begin{array}{c}a=5.06\\b=8.79\\c=9.75\\\beta = 100\end{array}$ & NA & NA\\
  \hline
c-BN         & $a=3.63$&4.61&2.34\\
 exp.\footnote{References \onlinecite{Eichhorn:bx0520,Levinshtein:2001wl}} &$a=3.616$&4.46&2.6\\
  \hline
h-BN&$\begin{array}{c}a=2.51\\c=6.66~\footnote{The lattice constant in the $c$-direction is fixed to the experimental value.}   \end{array}$&$\begin{array}{c} \epsilon_{\perp}=4.76\\\epsilon_{\parallel}=2.68 \end{array}$
    &$\begin{array}{c}\epsilon_{\perp}=1.83\\\epsilon_{\parallel}=0.44 \end{array}$\\
exp.\footnote{Reference \onlinecite{Levinshtein:2001wl}} &$\begin{array}{c}a=2.5\\c=6.66\end{array}$&$\begin{array}{c} \epsilon_{\perp}=4.3\\\epsilon_{\parallel}=2.2 \end{array}$
    &$\begin{array}{c}\epsilon_{\perp}=2.6\\\epsilon_{\parallel}=2.9 \end{array}$\\
  \hline
Si           &$a=5.47$&12.98& -\\
 exp.\footnote{References \onlinecite{Hubbard:a12589,Levinshtein:2001wl}} &$a=5.431$& 11.7 &-\\
  \hline
GaAs         & $a=5.63$&15.9&1.95\\
 exp.\footnote{References \onlinecite{Cooper:a03504,Komsa:2012ji}}  &$a=5.654$& 11.1& 2.0\\
  \hline
diamond     &$a=3.55$&5.58&-\\
 exp.\footnote{References \onlinecite{Hom:a12735,Komsa:2012ji}} &$a=3.567$&5.7&-\\
  \hline
  \hline

  \end{tabular}
 \end{threeparttable}
 \end{center}
 \label{list_properties}
\end{table}

\section{Image-charge correction}\label{image}
Here, we address the image-charge correction schemes that have been devised since a few decades ago.
In this study, we suppose that the defect charge is localized in the supercell.
Following Ref.~\onlinecite{Komsa:2012ji}, we consider three systems:
(1) a pristine bulk system, 
(2) a system with a periodic array of localized defects with charge $q$ and a neutralizing background charge with charge density $-\frac{q}{\Omega}$, where $\Omega$ is volume of the supercell, and
(3) a system with a single isolated defect with charge $q$.
The potential is represented with $V_{\rm bulk}$, $V_{{\rm defect},q}$, and $V_{{\rm isolated},q}$, respectively.
Here and hereafter, to avoid confusions, we preferentially adopt the signs based on conventional electrostatic potential following Ref.~\onlinecite{Komsa:2012ji}.
The electron charge is then set to the negative value.

Assume that charge density of a single defect within the supercell $\rho_d ({\bm r})$, which satisfies $q=\int_{\Omega} \rho_d ({\bm r}) d{\bm r}$, is the same in both periodic and isolated systems.
In other words, the variation of $\rho_d ({\bm r})$ induced by the spurious potential caused by the periodic images and background charge is negligibly small.
The electrostatic energy of a defect, its images, and the background charge of the periodic system is then written as
\begin{eqnarray}
 E_{\rm periodic} = \frac{1}{2} \int_{\Omega} \bigl(V_{{\rm defect},q} ({\bm r}) - V_{\rm bulk}({\bm r})\bigr) \biggr( \rho_d({\bm r}) -\frac{q}{\Omega} \biggl) d{\bm r}.
\end{eqnarray}
The factor $\frac{1}{2}$ accounts for removing double counting, and the integration is performed over the supercell.
The electrostatic energy of an isolated defect reads
\begin{eqnarray}
 E_{\rm isolated} = \frac{1}{2} \int \bigl(V_{{\rm isolated},q} ({\bm r}) - V_{\rm bulk}({\bm r})\bigr) \rho_d({\bm r}) d{\bm r}.
\end{eqnarray}
The integration is performed over entire space.
Following $\int_{\Omega} V_{{\rm defect},q} d{\bm r} = 0$ and $\int_{\Omega} V_{\rm bulk} d{\bm r} = 0$ by convention and the assumption that the defect charge is localized in the supercell,
the correction to the defect formation energy is written as~\cite{Dabo:2008eg,Komsa:2012ji}
\begin{eqnarray}
 E_{\rm cor} =  E_{\rm isolated} -  E_{\rm periodic} = \frac{1}{2} \int_{\Omega} V_{\rm cor} ({\bm r}) \rho_d({\bm r}) d{\bm r}
\label{correction_energy}
\end{eqnarray}
where $V_{\rm cor} = V_{{\rm isolated},q} - V_{{\rm defect},q}$.
This equation indicates that the image-charge correction is a {\it potential correction} for removing the spurious Coulomb potential caused by the defect images and background charge.
Note that although the background charge density is also removed via the correction, it does not contribute to the correction energy due to the convention of the zero average potential.

\subsection{Point-charge correction}
The simplest image-charge correction is to subtract the PC energy.
Only $V_{\rm cor}$ at the defect site is essential for the PC correction, and can be estimated by an Ewalt summation.
Fuchs derived the Ewalt formalism for the Madelung energy of periodically repeating PCs immersed
in a neutralizing background charge for the study of the stability of Cu metal.~\cite{Fuchs:1935}
Leslie and Gillan employed it for the correction of defect formation energies.~\cite{Leslie:1985bn}
Suppose that the long-range Coulomb interaction is screened by a macroscopic dielectric constant $\epsilon$ in the isotropic medium.
The potential at the defect site ${\bm R_0}$ caused by PCs with charge $q$ located at the periodic image sites ${\bm R_i}$~$(i \neq 0)$ and the background charge with charge density $-\frac{q}{\Omega}$, 
namely Madelung potential, can be written for a cubic cell as
\begin{eqnarray}
  V_{{\rm PC}, q}^{\rm iso} = - \frac{\alpha q}{\epsilon L} = \frac{q}{\epsilon} \Biggl\{ \sum_{\bm R_i}^{i \neq 0}\frac{{\rm erfc}(\gamma \left|{\bm R_i}\right|)}{\left|{\bm R_i}\right|} - \frac{\pi}{\Omega \gamma ^2} \nonumber \\ 
    + \sum_{\bm G_i}^{i \neq 0}\frac{4\pi}{\Omega} \frac{{\rm exp}(- {\bm G_i}^2 / 4\gamma^2)}{{\bm G_i}^2}  - \frac{2 \gamma}{\sqrt{\pi}} \Biggr\},
\label{Fuchs}
\end{eqnarray}
where the summation of ${\bm R_i}$ and ${\bm G_i}$ runs over all vectors of the direct and reciprocal lattices except ${\bm R_0}$ and ${\bm G_0}$ = ${\bm 0}$, 
and $L$ is the dimension of the supercell, $\alpha$ the Madelung constant which depends on the Bravais lattice, 
and $\gamma$ a suitably chosen convergence parameter which does not influence on the potential.~\cite{Fuchs:1935,Leslie:1985bn}
Here and hereafter, we suppose that a single defect exists in the supercell, and the basis is taken to be the defect site at ${\bm r}={\bm R_0}={\bm 0}$.
The second term, which is absent in the charge neutral Ewalt summation without the background charge,  
is essential for correcting the potential shift introduced by a periodic array of Gaussian charges instead of PCs in the third term,~\cite{Fuchs:1935} and obtained by
\begin{eqnarray}
  - \frac{1}{\Omega} \int_0^{\infty} \frac{{\rm erfc}(\gamma r)}{r} \cdot 4 \pi r^2 dr = - \frac{\pi}{\Omega \gamma ^2}.
\end{eqnarray}
The forth term corresponds to the cancellation of the potential introduced by the Gaussian located at ${\bm r} = {\bm 0}$ which is included in the third term.
The correction potential is then $V_{\rm cor} = - V_{{\rm PC}, q}^{\rm iso}$.
This is of course the same as the functional derivative of the PC correction energy with respect to the defect charge density.~\cite{Taylor:2011ue, Komsa:2012ji}
The PC correction energy is then written as 
\begin{eqnarray}
E_{\rm PC}^{\rm iso} = \frac{1}{2} \int_{\Omega} (- V^{\rm iso}_{{\rm PC},q}) \cdot q \delta({\bm r}) d{\bm r} = - \frac{q}{2}V_{{\rm PC},q}^{\rm iso}=\frac{q^2 \alpha}{2L}.
\end{eqnarray}

Strictly, the use of a dielectric constant is valid only for cubic systems, and it must be replaced by a dielectric tensor $\overline\epsilon$ for the others.
This extension is promising for layered and low-dimensional materials such as nanowires and nanosheets.~\cite{Rurali:2009io,Murphy:2013da}
The Madelung potential in Eq.~(\ref{Fuchs}) is then rewritten as~\cite{Rurali:2009io,Murphy:2013da} 
\begin{eqnarray}
  V_{{\rm PC},q}^{\rm aniso} &=& \sum_{\bm R_i}^{i \neq 0}\frac{q}{\sqrt{\left|\overline{\epsilon}\right|}} 
      \frac{{\rm erfc}(\gamma \sqrt{ {\bm R_i} \cdot \overline\epsilon^{-1} \cdot {\bm R_i} })}{\sqrt{ {\bm R_i} \cdot \overline\epsilon^{-1} \cdot {\bm R_i} }}
     -  \frac{\pi q}{\Omega \gamma ^2} \nonumber \\
    &+& \sum_{\bm G_i}^{i \neq 0}\frac{4\pi q}{\Omega}
  \frac{{\rm exp}(- {\bm G_i} \cdot \overline\epsilon \cdot {\bm G_i} / 4\gamma^2)}{{\bm G_i} \cdot \overline\epsilon \cdot {\bm G_i}} - \frac{2\gamma q}{\sqrt{\pi \left|\overline{\epsilon}\right|}}.
\label{anisotropic_pc}
\end{eqnarray}
The correction energy is written as $E_{\rm PC}^{\rm aniso}= - \frac{q}{2}V_{{\rm PC},q}^{\rm aniso}$.
Rurali and Cartoix$\grave{\rm a}$ calculated the Al substitution energy with this correction in one-dimensional Si nanowire,~\cite{Rurali:2009io}
and Murphy and Hine corrected the formation energies of Ti vacancy ($V_{\rm Ti}^{-4}$), Li antisite on Ti (${\rm Li}_{\rm Ti}^{-3}$), and oxygen interstitial (${\rm O}_i^{-2}$)
in monoclinic $\beta$-Li$_2$TiO$_3$.~\cite{Murphy:2013da}

\subsection{Makov-Payne correction}
The PC correction is the leading term of the image-charge correction with the $L^{-1}$ order.
Makov and Payne (MP) then derived the correction term with the $L^{-3}$ order.~\cite{Makov:1995vk}
Dabo {\it et al}. also derived the same formula in a simpler and physically intuitive manner.~\cite{Dabo:2008eg}
Following Refs.~\onlinecite{Makov:1995vk} and \onlinecite{Dabo:2008eg}, the correction potential $V_{\rm cor}$ for a defect in a periodically repeated cubic cell is written as
\begin{eqnarray}
  V_{\rm MP}^{\rm iso}({\bm r}) =  -V_{\rm PC,q}^{\rm iso} - \frac{2 \pi q}{3\epsilon L^3}r^2 + \frac{4 \pi}{3\epsilon L^3} {\bm p} \cdot {\bm r} - \frac{2 \pi Q}{3\epsilon L^3} + O(r^4).
 \label{Dabo}
\end{eqnarray}
Here, ${\bm p}=\int {\bm r} \rho_d ({\bm r}) d{\bm r}$ is dipole moment and $Q=\int r^2 \rho_d ({\bm r}) d{\bm r}$ second radial moment.
The correction energy under the cubic symmetry up to the $L^{-3}$ order is then
\begin{eqnarray}
  E_{\rm MP}^{\rm iso} = \frac{1}{2} \int_{\Omega}V_{\rm MP}^{\rm iso}({\bm r}) \rho_d({\bm r}) d{\bm r} = E_{\rm PC}^{\rm iso} - \frac{2 \pi qQ}{3\epsilon L^3} + \frac{2 \pi {\bm p}^2}{3\epsilon L^3} + O(L^{-5}).
 \label{Makov-Payne}
\end{eqnarray}
Assuming that the dipole moment is negligible, the third term is omitted.
For charged ions and molecules in vacuum under periodic boundary conditions, we can exactly calculate $E^{\rm iso}_{\rm MP}$ up to the third order as discussed in Sec.~\ref{alignment_revisited}.
However, there are some problems for defects in crystalline materials.
Firstly, the defect charge $\rho_d$ is ill defined, because the immersed $\rho_d$ and screening charge are inseparable; thus $Q$ cannot be calculated directly.~\cite{Lany:2008gk,Taylor:2011ue,Komsa:2012ji}
Secondly, the Coulomb interaction is assumed to be screened by a macroscopic dielectric constant, which is correct only for cubic systems.
It is also doubtful that the short-range Coulomb interaction is assumed to be screened by the macroscopic dielectric constant.
Therefore, $E^{\rm iso}_{\rm MP}$ is usually not applied and the correction energy is determined by fitting the energies calculated with various supercells 
with different sizes and shapes.~\cite{PhysRevB.77.245202,PhysRevB.73.035215}
Such calculations need plently of computational costs especially for larger supercells, to which advanced DFT and many-body theory calculations are not accessible.

\subsection{FNV correction}
Later on, Freysoldt, Neugebauer, and Van de Walle proposed a novel correction scheme.~\cite{Freysoldt:2009ih}
Our main purpose in this study is to extend this scheme to be applied to broad classes of materials and assess its performance.
Following Refs.~\onlinecite{Freysoldt:2011hf} and \onlinecite{PhysRevB.88.115104}, the correction energy of the FNV scheme is expressed as
\begin{eqnarray}
  E_{\rm FNV} =  E_{\rm PC} - q \Delta V_{{\rm PC}, {q/b}}|_{\rm far}.
 \label{FNV}
\end{eqnarray}
$\Delta V_{{\rm PC}, {q/b}}$ is the potential difference between the defect-induced potential
\begin{eqnarray}
V_{q/b} = V_{{\rm defect},q} - V_{\rm bulk},
 \label{defect_potential}
\end{eqnarray}
 and the PC potential $V_{{\rm PC}, q}$,~\cite{Freysoldt:2009ih,Freysoldt:2011hf,Komsa:2012ji,PhysRevB.88.115104}
\begin{eqnarray}
  \Delta V_{{\rm PC}, {q/b}} = V_{q/b} - V_{{\rm PC}, q}.
\label{far}
\end{eqnarray}
$\Delta V_{{\rm PC}, {q/b}}|_{\rm far}$ is $\Delta V_{{\rm PC}, {q/b}}$ at a place far from the defect in the supercell.
Instead of a Gaussian charge originally adopted in Ref.~\onlinecite{Freysoldt:2009ih} as a model charge for the localized defect in the supercell, we use a PC.
This is because the PC model can be readily rewritten in the anisotropic form, 
and the correction energy can be divided into physically-meaningful long-range Coulomb interaction part and short-range part.~\cite{PhysRevB.88.115104}
The latter can also be attained with Gaussian by redefining the long-range Coulomb interaction energy and alignmentlike term.~\cite{Freysoldt:2011hf}

\begin{figure*}
  \includegraphics[width=1\linewidth]{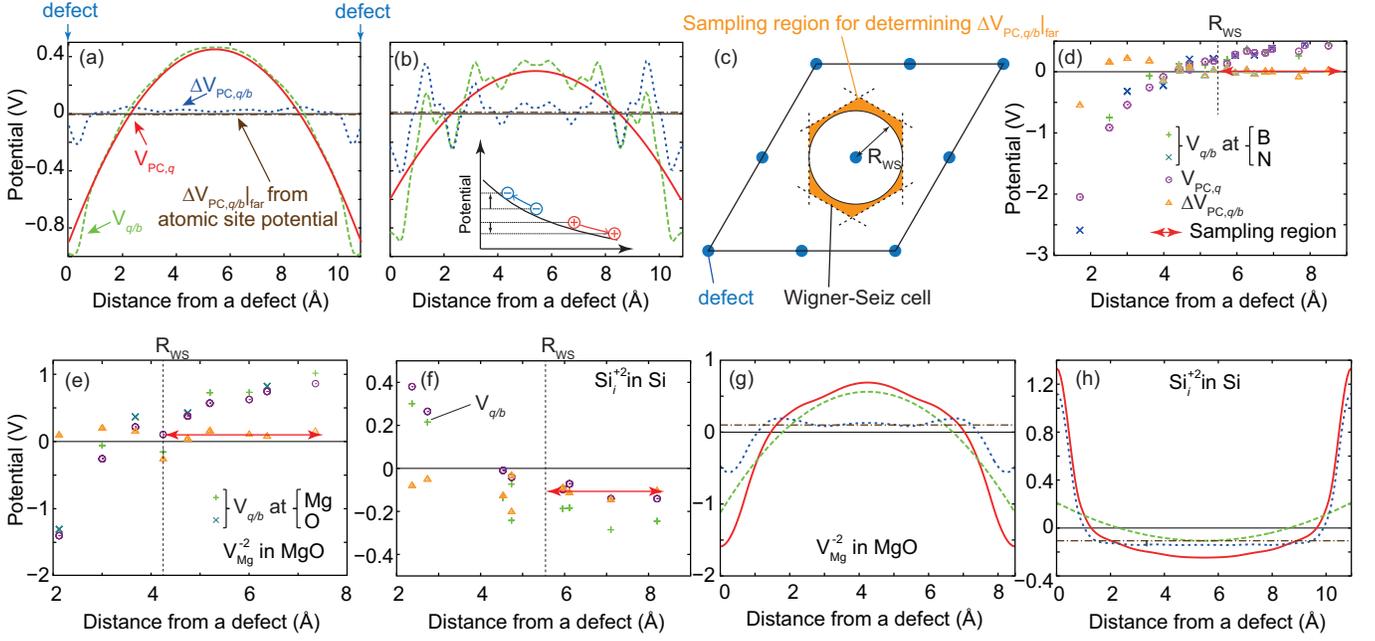}
  \caption{(a--b) Planar-averaged defect-induced potential $V_{q/b}$, PC potential $V_{{\rm PC},q}$, and their difference $\Delta V_{{\rm PC}, {q/b}}$ 
of (a) unrelaxed and (b) relaxed $V_{\rm B}^{-3}$ in c-BN with a $3\times3\times3$ supercell containing 215 atoms.
The $\Delta V_{{\rm PC},{q/b}}|_{\rm far}$ obtained using atomic site potential is also depicted for comparison.
Inset in (b): Schematic illustration showing the displacements of polarizable ions under electric field caused by the charged defect.
(c) Schematic of sampling region used for estimating $\Delta V_{{\rm PC}, {q/b}}|_{\rm far}$ by averaging atomic site $\Delta V_{{\rm PC}, {q/b}}$ 
at the region outside of the sphere in contact with the Wigner-Seiz cell.
Note that the sampling region depends only on the Bravais lattice of the supercell.
(d) $V_{q/b}$, $V_{{\rm PC},q}$, and $\Delta V_{{\rm PC},{q/b}}$ at the atomic positions in the supercell of the relaxed $V_{\rm B}^{-3}$ in c-BN.
The region for averaging $\Delta V_{{\rm PC},{q/b}}$ and its averaged value are expressed in the width and height of the arrow, respectively.
(e, f) $V_{q/b}$, $V_{{\rm PC},q}$, and $\Delta V_{{\rm PC},{q/b}}$ at the atomic sites of the unrelaxed (e) $V_{\rm Mg}^{-2}$ in MgO and (f) Si$_i^{+2}$ in Si with $2\times2\times2$ supercells 
constructed from the conventional unit cells.
(g, h) Planar-averaged $V_{q/b}$, $V_{{\rm PC},q}$, and $\Delta V_{{\rm PC},{q/b}}$ of the unrelaxed (g) $V_{\rm Mg}^{-2}$ and (h) Si$_i^{+2}$
together with $\Delta V_{{\rm PC},{q/b}}|_{\rm far}$ in (e) and (f).}
  \label{alignment}
\end{figure*}

The second term in Eq.~(\ref{FNV}) is denoted as potential alignmentlike term.~\cite{Freysoldt:2009ih,Komsa:2012ji}
An important point is that this alignmentlike term is different from the conventional potential alignment correction
and approximately corresponds to the MP third order term when the PC model is used.~\cite{Komsa:2012ji,PhysRevB.88.115104}
When $\rho_d$ has spherical distribution, 
the defect-induced potential outside of the defect coincides with the PC potential under the open boundary condition, whereas they are different under the periodic boundary conditions. 
This discrepancy is due to the convention that the potential average in the entire supercell is set to zero.
Komsa {\it et al.} have discussed this point in detail in Ref.~\onlinecite{Komsa:2012ji} and derived the relationship 
$\Delta V_{{\rm PC}, {q/b}}|_{\rm far} = \frac{2 \pi Q}{3\epsilon \Omega}$ in an isotropic medium.
This spurious potential shift caused by the periodic boundary condition has to be removed for charged defects, 
and its correction corresponds to the alignmentlike term.
The great advantage of the FNV scheme is that we do not have to know the details of microscopic screening 
and their coupling to the actual unscreened or partially screened charge distribution beyond the PC model because these effects are incorporated into the alignmentlike term. 
Another advantage is that any shapes of supercells are applicable as long as the defect charge is encased in the supercell.

Although it is originally proposed to use either neutral defect or pristine bulk for a reference potential for estimating $\Delta V_{{\rm PC}, {q/b}}$, we use the pristine bulk only.
This is because the defect-induced potential can be quantified as a variation of the potential relative to the pristine host,
and there is no reason that a system with a neutral defect can be used as a reference.
Especially, the alignmentlike term estimated with the neutral defect with delocalized carriers is erroneous. 
Komsa {\it et al.} proposed a way to estimate $\Delta V_{{\rm PC}, {q/b}}$ by using potential of a neutral defect system as a reference, 
and perform the conventional potential alignment between the neutral defect and pristine bulk systems.~\cite{Komsa:2012ji}
Their approach is conceptually different but the correction energy is the same as ours.

\subsection{Application of atomic site potential as a potential marker}
Originally the FNV scheme uses planar-averaged electrostatic potential for determining $\Delta V_{{\rm PC},{q/b}}|_{\rm far}$.~\cite{Freysoldt:2009ih}
This, however, does not work properly when geometry optimization is performed.
It is especially significant for an ionic host, in which long-range Coulomb interaction is screened by the dipoles of polarizable ions.
This is demonstrated in Figs.~\ref{alignment} (a) and (b) that show the planar-averaged defect-induced potential $V_{q/b}$, PC potential $V_{{\rm PC}, q}$, 
and their difference $\Delta V_{{\rm PC}, {q/b}}$ for the unrelaxed and relaxed B vacancy in the -3 charge state ($V_{\rm B}^{-3}$) in c-BN.
In the unrelaxed geometry, both $V_{q/b}$ and $V_{{\rm PC}, q}$ show parabolic shape far from the defect, which comes from the homogeneous background charge through the Poisson's equation.
Their difference then reaches a plateau between the defect and its periodic image, and $\Delta V_{{\rm PC},{q/b}}|_{\rm far}$ can be defined with small uncertainty.
On the other hand, $V_{q/b}$ becomes scraggly in the relaxed geometry reflecting the atomic displacements, whereas $V_{{\rm PC}, q}$ remains parabolic.
As a result $\Delta V^{\rm PC}_{q/b}|_{\rm far}$ cannot be determined properly.

An alternative way is to employ atomic site electrostatic potential.
This is often utilized for the potential alignment in defect calculations~\cite{Lany:2008gk,Choi:2013uo} as well as the
determination of ionization potential and band offsets in semiconductors and insulators.~\cite{PhysRevB.88.035305,PhysRevB.86.245433,Gruneis.2014}
Screened potential at the arbitrary position ${\bm r} \neq {\bm 0}$ in an anisotropic dielectric medium reads
\begin{eqnarray}
  V_{{\rm PC},q}^{\rm aniso}({\bm r} \neq {\bm 0}) = \sum_{\bm R_i}\frac{q}{\sqrt{\left|\overline{\epsilon}\right|}} 
      \frac{{\rm erfc}(\gamma \sqrt{ {\bm R_i} \cdot \overline\epsilon^{-1} \cdot {\bm R_i} })}{\sqrt{ {\bm R_i} \cdot \overline\epsilon^{-1} \cdot {\bm R_i} }}
     -  \frac{\pi q}{V_c \gamma ^2} \nonumber \\
    + \sum_{\bm G_i}^{i \neq 0}\frac{4\pi q}{V_c}
      \frac{{\rm exp}(- {\bm G_i} \cdot \overline\epsilon \cdot {\bm G_i} / 4\gamma^2)}{{\bm G_i} \cdot \overline\epsilon \cdot {\bm G_i}} \cdot {\rm exp}(i{\bm G_i} \cdot {\bm r}).
\label{ani_pc_potential}
\end{eqnarray}
This is used for evaluating $\Delta V_{{\rm PC},{q/b}}$ in Eq.~(\ref{far}).
We should keep in mind that the farthest atomic site from the defect is not necessarily the best reference for evaluating $\Delta V_{{\rm PC},{q/b}}|_{\rm far}$.
This is because (i) the farthest atom lies between the defect and its periodic image, and might be suffered from an artificial defect-defect interaction in smaller supercells, and
(ii) the displacements of the polarizable ions as a result of the screening may bias the electrostatic potential as illustrated in the inset of Fig.~\ref{alignment}(b).
Thus, we instead propose to average $\Delta V_{{\rm PC},{q/b}}|_{\rm far}$ at the atomic positions in the region outside of the sphere that is in contact with the Wigner-Seiz cell with radius $R_{WS}$
as illustrated in Fig.~\ref{alignment}(c).
We call this region sampling region.
This averaging is justified by the assumption that the defect charge spherically distributes and is encased in the supercell.
It is also advantageous that the sampling region does not depend on the choice of the supercell as long as the Bravais lattice is same.
As an example, the atomic site $V_{q/b}$, $V_{{\rm PC},q}$, and $\Delta V_{{\rm PC},{q/b}}$ of $V_{\rm B}^{-3}$ in c-BN are shown in Fig.~\ref{alignment}(d).
$\Delta V_{{\rm PC},{q/b}}$ shows scattering behavior near the defect, but it converges at the outside of $R_{WS}$.

A disadvantage of the use of the atomic site potential is that the number of atomic sites for determining $\Delta V_{{\rm PC},{q/b}}|_{\rm far}$ might not be sufficient in small supercells, 
and non-negligible sampling errors might be involved.
To check the accuracy, we compare the averaged atomic site $\Delta V_{{\rm PC},{q/b}}|_{\rm far}$ with planar-averaged $\Delta V_{{\rm PC},{q/b}}|_{\rm far}$.
In Figs.~\ref{alignment}(e--h), we show atomic site and planar-averaged $V_{q/b}$, $V_{{\rm PC},q}$, 
and $\Delta V_{{\rm PC},{q/b}}$ of Mg vacancy ($V_{\rm Mg}^{-2}$) in MgO and Si self-interstitial at the tetrahedral site (Si$_i^{+2}$) in Si.
For comparison we used relatively small $2\times 2\times 2$ supercells constructed from the conventional unit cells and did not relax the atomic positions.
Between the defect and its image, the planar-averaged $\Delta V_{{\rm PC},{q/b}}$ almost converge in both defect systems, indicating the defect charge is well localized in the supercells.
$\Delta V_{{\rm PC},{q/b}}|_{\rm far}$ determined from the atomic site potential at the sampling region are almost the same for $V_{\rm Mg}^{-2}$ in MgO and Si$_i^{+2}$ in Si; 
the differences are less than 40 meV in both systems.
Note that $\Delta V_{{\rm PC},{q/b}}$ at the farther atomic site is almost same as the averaged $\Delta V_{{\rm PC},{q/b}}|_{\rm far}$.
When the cell size increases, these differences and consequently sampling errors drastically reduce, owing to the increase of the sampling points for evaluating $\Delta V_{{\rm PC},{q/b}}$.

\begin{figure*}
  \includegraphics[width=1\linewidth]{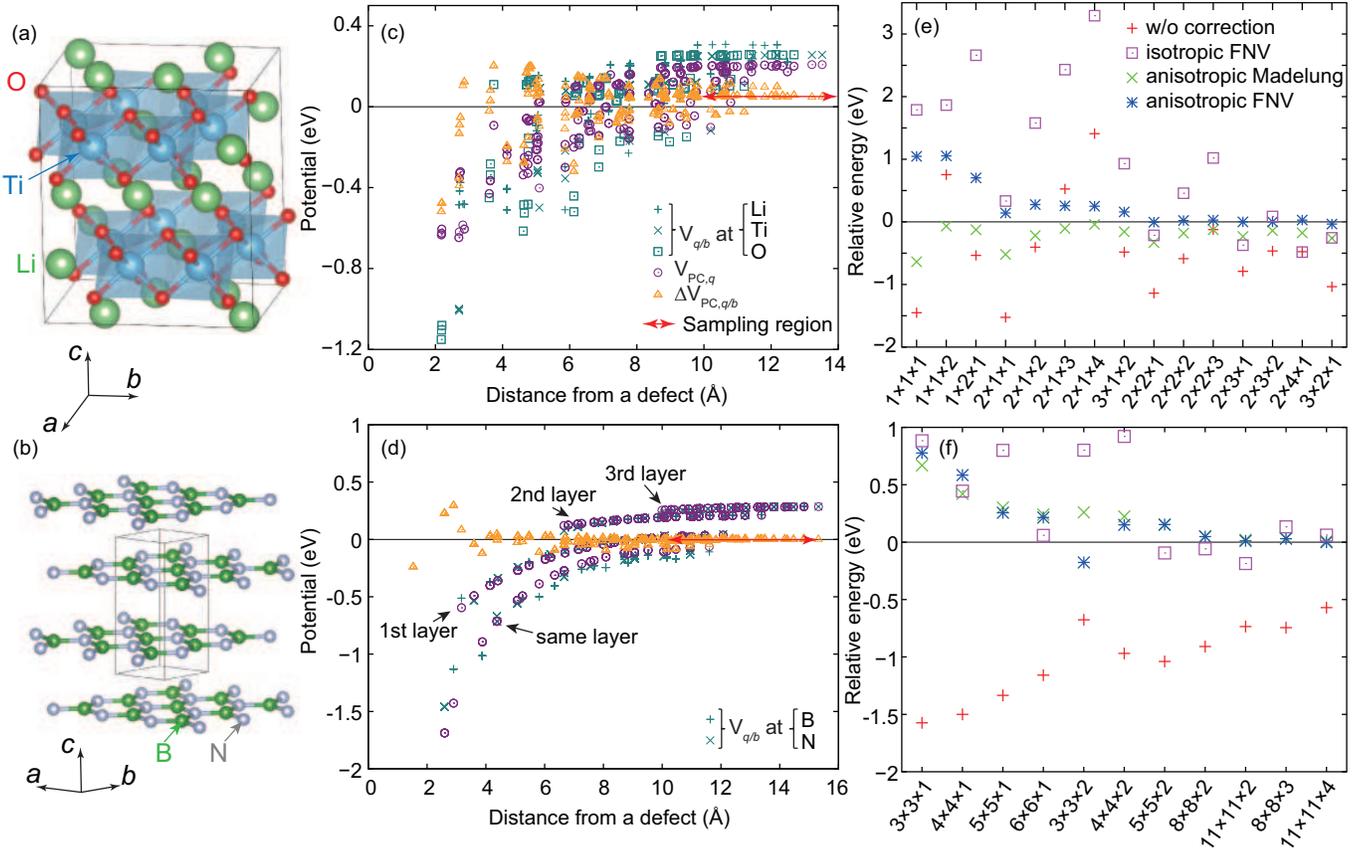}
  \caption{(a--b) Crystal structures of (a) $\beta$-Li$_2$TiO$_3$ and (b) h-BN.
    The unit cells of $\beta$-Li$_2$TiO$_3$ and h-BN contain 48 and 4 atoms, respectively.
    (c--d) $V_{q/b}$, $V_{{\rm PC},q}$, and $\Delta V^{\rm PC}_{q/b}$ at the atomic sites in 
    (c) $V_{\rm Ti}^{-4}$ in the $2\times 2\times 2$ supercell of $\beta$-Li$_2$TiO$_3$ (383 atoms) and (d) B$_{\rm N}^{+2}$ in the $8\times 8\times 3$ supercell of h-BN (768 atoms).
    (e--f) Relative formation energies of (e) $V_{\rm Ti}^{-4}$ and (f) B$_{\rm N}^{+2}$ as a function of the supercell size and shape.
    Zeros are set to the formation energies calculated with the largest supercells and anisotropic FNV corrections.
    Atomic relaxations are considered in any cases.}
  \label{anisotropic}
\end{figure*}

\subsection{Assessment of the performance of the extended FNV scheme}
Here, we discuss the performance of the extended FNV scheme using the anisotropic PC model.
The test calculations were performed for the Ti vacancy ($V_{\rm Ti}^{-4}$) in $\beta$-Li$_2$TiO$_3$ and B antisite defect on N (B$_{\rm N}^{+2}$) in h-BN.
Their crystal structures are shown in Figs.~\ref{anisotropic}(a) and (b).
As can be inferred from the layered structures, the dielectric tensors have very different diagonal components as listed in Table ~\ref{list_properties}.

Figures~\ref{anisotropic}(c) and (d) show the atomic site $V_{q/b}$, $V_{{\rm PC}, q}$, and $\Delta V_{{\rm PC}, {q/b}}$
of $V_{\rm Ti}^{-4}$ in the  $\beta$-Li$_2$TiO$_3$ $2\times 2\times 2$ supercell and B$_{\rm N}^{+2}$ in the h-BN $8\times 8\times 3$ supercell.
$V_{q/b}$ widely scatter even at the same distance from the defect, reflecting the anisotropic screening feature.
Interestingly, $V_{q/b}$ in B$_{\rm N}^{+2}$ can be clearly divided into layer-by-layer components.
$\Delta V_{{\rm PC}, {q/b}}$ in B$_{\rm N}^{+2}$ is almost constant except the immediate vicinity of the defect, indicating that the defect charge is very localized.
On the other hand, $\Delta V_{{\rm PC}, {q/b}}$ of $V_{\rm Ti}^{-4}$ in $\beta$-Li$_2$TiO$_3$ is widespread and converges in a region far from the defect.

We corrected their formation energies based on Eq.~(\ref{FNV}).
$E_f[V_{\rm Ti}^{-4}]$ in $\beta$-Li$_2$TiO$_3$ and $E_f[{\rm B}_{\rm N}^{+2}]$ in h-BN without corrections, with FNV corrections in the isotropic form,
where the average of the diagonal components of the dielectric tensor is used as a dielectric constant $\epsilon = \langle \epsilon_{ii} \rangle$, 
and PC and FNV corrections in the anisotropic form are plotted in Figs.~\ref{anisotropic}(e) and (f) for a range of supercell sizes and shapes.
As discussed later, the potential alignment is not considered for avoiding double counting of the correction term.

Without corrections, $E_f[V_{\rm Ti}^{-4}]$ in $\beta$-Li$_2$TiO$_3$ widely scatters depending on the supercell size and shape.
The isotropic FNV correction with a dielectric constant, which is a typical approximation, 
does not avail to correct $E_f[V_{\rm Ti}^{-4}]$; in the elongated supecells, it makes $E_f[V_{\rm Ti}^{-4}]$ even worse.
On the other hand, the anisotropic PC drastically reduces the cell size/shape dependence of $E_f[V_{\rm Ti}^{-4}]$ as also reported in Ref.~\onlinecite{Murphy:2013da}.
The potential alignmentlike term in the anisotropic FNV scheme corrects the remaining cell size/shape dependence, and it almost vanishes in large supercells.
As a result we see the extension along $a$-axis is essential for accurate estimation of $E_f[V_{\rm Ti}^{-4}]$, and the $2\times 1\times 1$ 95-atom supercell would be a good compromise
for the computationally expensive first-principles calculations such as hybrid functional calculations.
Similarly, the anisotropic PC correction significantly improves $E_f[{\rm B}_{\rm N}^{+2}]$ in h-BN, but the alignmentlike term is quite small in this case.
$E_f[{\rm B}_{\rm N}^{+2}]$ is systematically overestimated when the $c$-axis is not expanded in the supercell.
In this case, BN sheets with and without defects alternate layer-by-layer, 
and it would not be appropriate to use a static dielectric constant $\epsilon^{\rm ele} + \epsilon^{\rm ion}$ along $c$-direction.
Thus, good compromise for $E_f[{\rm B}_{\rm N}^{+2}]$ would be the $4\times4\times2$ 128-atom supercell, which is expected to have an error less than 0.15 eV.

\section{Potential alignment revisited}\label{alignment_revisited}
As mentioned above, there is a longstanding controversy over the potential alignment.
We here demonstrate that the potential alignment is not needed when the image-charge correction is applied properly.
Indeed, some authors refrain from adopting both potential alignment {\it and} image-charge corrections because it might include a part of double counting terms.~\cite{PhysRevB.77.245202,Chen:2010ea,Taylor:2011ue}
As indicated in Eq.~(\ref{correction_energy}), image-charge correction is a {\it potential correction}, and it changes the potential $V_{{\rm defect},q}$ to $V_{{\rm isolated},q}$.
Then, $V_{{\rm isolated},q} = V_{{\rm defect},q} + V_{\rm cor} = V_{\rm bulk} +V_{q/b} + V_{\rm cor}$, and $V_{q/b} + V_{\rm cor}$ is the potential induced by a single defect.
The proper potential alignment is achieved at the point infinitely far from the defect, and $\displaystyle \lim_{\left|{\bm r}\right| \to \infty}(V_{q/b} + V_{\rm cor}) = 0$.
Hence, after adopting the image-charge correction, the potential of the supercell with a single defect is aligned to the bulk potential, indicating that the potential alignment is unnecessary for estimating the charged defect formation energy.

\begin{figure*}
  \includegraphics[width=1\linewidth]{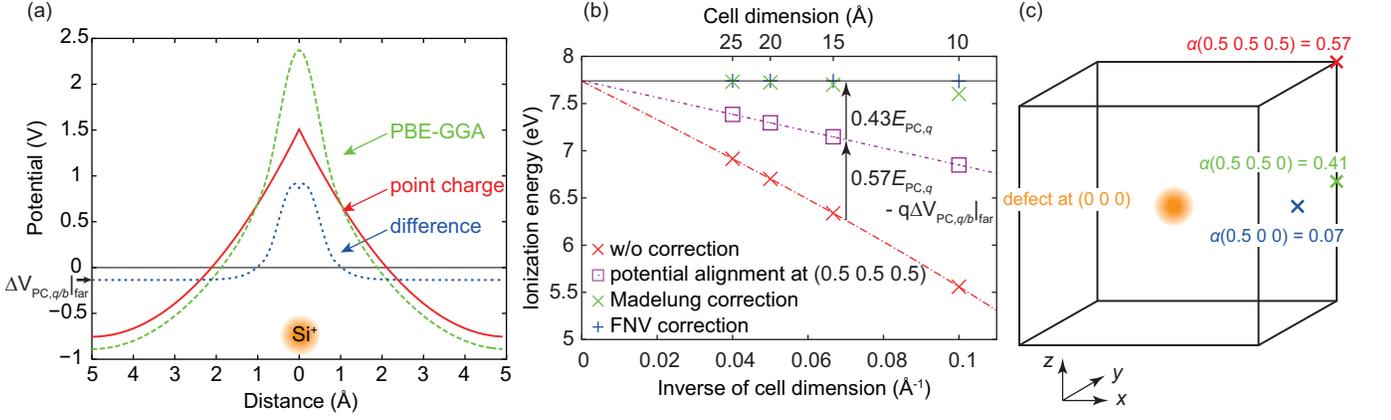}
  \caption{
(a) Planar-averaged electrostatic potential of a Si$^{+}$ ion calculated with 
the PBE-GGA located in a $10~\AA \times 10~\AA \times 10~\AA$ cell, the PC potential with the +1 charge, and their difference.
The potential difference converges to $\Delta V_{{\rm PC},{q/b}}|_{\rm far} = \frac{2 \pi Q}{3L^3} (<0)$.
(b) Cell size dependence of the uncorrected and corrected ionization energy of a Si atom, $E({\rm Si}^{+}~{\rm ion}) - E({\rm Si}~{\rm atom}) + E_{\rm cor}$.
The ionization energy becomes almost independent of the cell dimension after applying both the PC correction and alignmentlike correction (the FNV correction).
(c) The fractions $\alpha$ of the PC correction implicitly included in the potential alignment at three points written in fractional coordinates in the cubic cell in an isotropic medium. 
Note that the alignmentlike term is fully included in the potential alignment at any point (see text).
}
  \label{point_charge_model}
\end{figure*}

The situation is analogous to the isolated charged ion in the cell under periodic boundary conditions, where the reference is not the pristine bulk but vacuum.
In this case, the total energy depends on the undetermined shift of the eigenvalues, and thus the potential at vacuum must be aligned to be zero.
Because the screening charge is absent and the ionic charge $\rho_{\rm ion}$ is well definied, $Q=\int r^2 \rho_{\rm ion} ({\bm r}) d{\bm r}$ is calculated exactly.
Figure~\ref{point_charge_model}(a) shows the planar-averaged electrostatic potential of a Si$^{+}$ ion 
obtained from a selfconsistent calculation with the PBE-GGA and of the PC model with the +1 charge, 
and their difference in a $10~\AA \times 10~\AA \times 10~\AA$ cell.
Around the Si$^{+}$ ion, the electrostatic potential is substantially different from the PC potential because of the finite distribution of the electrons.
It, however, becomes almost parallel to the PC potential at a distance of $\sim2~\AA$ from the ion, and the difference converges to a constant of $\frac{2 \pi Q}{3L^3}$.
Therefore, after applying the PC correction and alignmentlike correction, i.e. the FNV correction
the electrostatic potential far from the Si$^{+}$ ion changes to zero, meaning that the outermost potential from the Si$^{+}$ ion is already aligned to zero.
Note that the alignmentlike correction is almost the same as the MP third order correction via an explicit calculation of $Q$ 
(the difference is 0.02 meV in $10~\AA \times 10~\AA \times 10~\AA$ cell) because no screening exists.
Figure~\ref{point_charge_model}(b) shows the cell size dependence of the ionization energy of the Si atom.
One can see that the ionization energy with the FNV correction (sum of the PC correction and potential alignmentlike correction) does not show the cell size dependence, 
indicating the unnecessity of the additional potential alignment from energetics viewpoint.

We should emphasize that when the potential alignment is performed at a particular atomic site {\it before} the image-charge correction, 
a part of the PC correction is included in addition to the alignmentlike term.
This can be understood by writing the potential alignment term as 
\begin{eqnarray}
-q V_{q/b}({\bm r}) &=& -q (V_{{\rm PC},q}^{\rm iso}({\bm r}) + \Delta V_{{\rm PC}, {q/b}}|_{\rm far}) \nonumber \\
                   &=& \alpha ({\bm r})E_{\rm PC}^{\rm iso} -q \Delta V_{{\rm PC}, {q/b}}|_{\rm far}
\end{eqnarray}
with Eq.~(\ref{far}), where the potential alignment is performed at ${\bm r}$ outside of the defect.
Fractions $\alpha$ of the PC correction included in the potential alignment are calculated from $\alpha ({\bm r}) = -q V_{{\rm PC},q}^{\rm iso}({\bm r}) / E_{\rm PC}^{\rm iso} $ in an isotropic medium.
Note that $\alpha$ depends only on the fractional coordinates and supercell shape.
Figure~\ref{point_charge_model}(c) shows $\alpha$ at (0.5 0 0), (0.5 0.5 0), and (0.5 0.5 0.5)  in fractional coordinates in cubic systems.
For instance, when the potential at (0.5 0.5 0.5) is aligned to the bulk potential, 57~\% of the PC correction and 100~\% of the alignmentlike term are incorporated.
This is demonstrated for the Si ionization energy.
Figure~\ref{point_charge_model}(b) shows the corrected ionization energies by the potential alignment at (0.5 0.5 0.5).
They have cell size dependence linear to $L^{-1}$, and the rest of the correction energy corresponds to 43~\% of the PC correction.
41~\% and 7~\% of PC correction are included if the potential alignment is made at (0.5 0.5 0) and (0.5 0 0), respectively.

Lany and Zunger have reported that no significant third-order contribution of image-charge correction remains for the As vacancy with the +3 charge ($V_{\rm As}^{+3}$) in GaAs 
after the potential alignment.~\cite{Lany:2008gk,Lany:2009vf}
They explained it by calculating the second radial moment in the MP third order term using the total charge density difference between the charged and neutral DFT calculations.
However, their explanation leads to some conceptual difficulties as pointed by Komsa {\it et al}~\cite{Komsa:2012ji} and Lambrecht.~\cite{Lambrecht:2010jt}
The FNV correction energy can be rewritten as
\begin{eqnarray}
  E_{\rm FNV} &=&  E_{\rm PC} - q \Delta V_{{\rm PC}, {q/b}}|_{\rm far} \nonumber \\
             &=&  (1 - \alpha({\bm r})) E_{\rm PC} - q V_{q/b}({\bm r}),
\label{FNV_PA}
\end{eqnarray}
and $(1 - \alpha({\bm r})) E_{\rm PC}$ has $L^{-1}$ dependence as long as the potential alignment is attained at the same fractional coordinates in the supercells with the same shape.
Although in Ref.~\onlinecite{Lany:2009vf} the potential alignment was achieved by averaging the potential offset at atomic sites except for the immediate neighbors of the defect
and therefore $\alpha$ is unclear, we believe the absence of the third-order contribution is explained with Eq.~(\ref{FNV_PA}).
Our results support this as shown in the next section.

\section{Applications to defects in diverse materials}\label{test}
\begin{figure*}
  \includegraphics[width=1\linewidth]{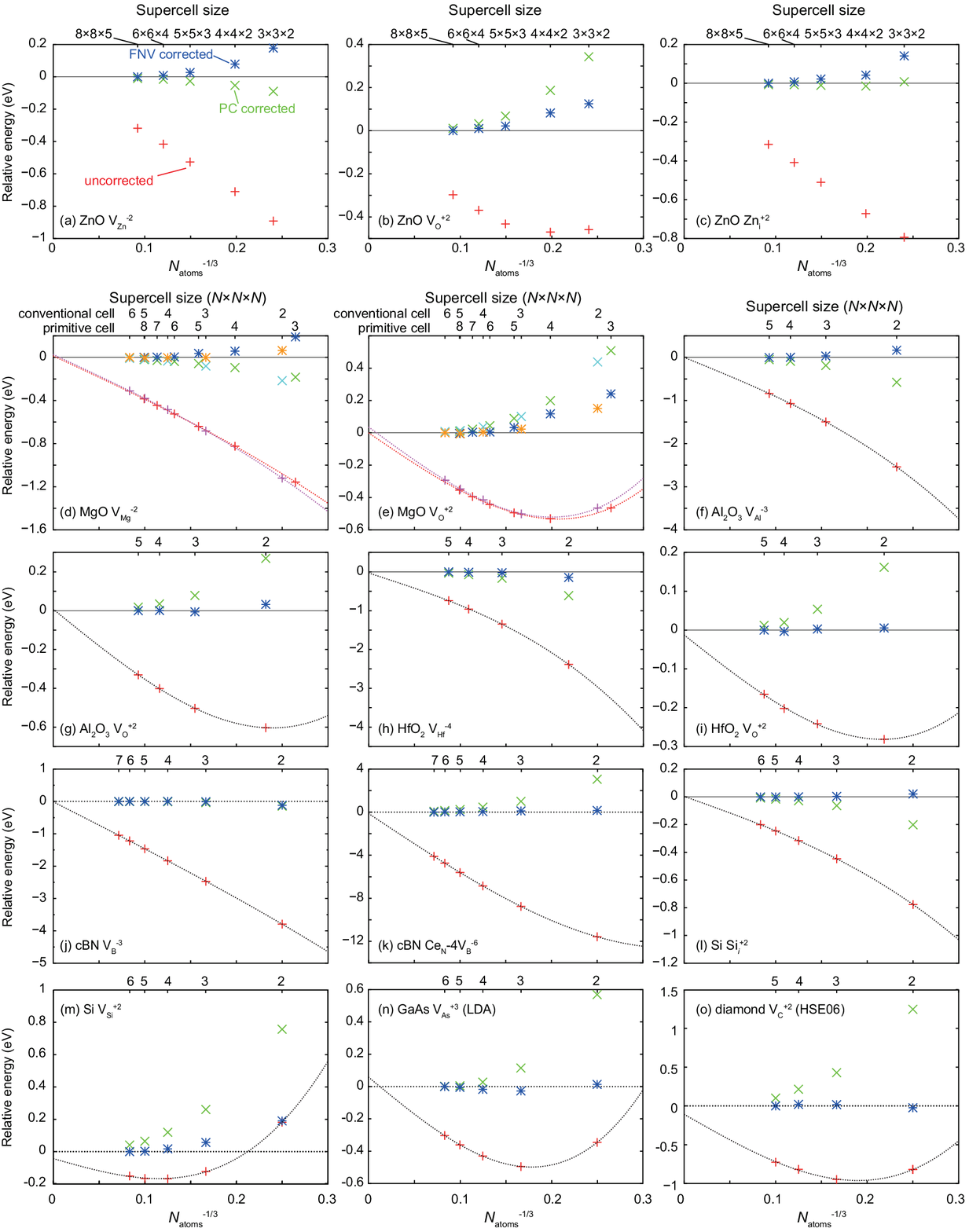}
  \caption{Relative formation energies of
           (a) $V_{\rm Zn}^{-2}$, (b) $V_{\rm O}^{+2}$, and (c) Zn$_i^{+2}$ in ZnO, (d) $V_{\rm Mg}^{-2}$ and (e) $V_{\rm O}^{+2}$ in MgO, (f) $V_{\rm Al}^{-3}$ and (g) $V_{\rm O}^{+2}$ in Al$_2$O$_3$, 
           (h) $V_{\rm Hf}^{-4}$ and (i) $V_{\rm O}^{+2}$ in HfO$_2$, (j) $V_{\rm B}^{-3}$ and (k) Ce$_{\rm N}$-4$V_{\rm B}^{-6}$ in c-BN, and (l) Si$_i^{+2}$ and (m) $V_{\rm Si}^{+2}$ in Si, 
          (n) $V_{\rm As}^{+3}$ in GaAs, and (o) $V_{\rm C}^{+2}$ in diamond with atomic relaxation considered.
    Zeros are set to the anisotropic FNV corrected defect formation energies calculated with the largest supercells.
    The horizontal axis is shown as a function of inverse of the cube root of the number of atoms.
    In cases where the cell dimension is isotropically expanded, the uncorrected defect formation energies are fitted with a function of $aN_{\rm atoms}^{-1}+bN_{\rm atoms}^{-1/3}+c$.
 }
  \label{list_defect_formation_energy}
\end{figure*}

To assess the accuracy of the extended FNV scheme, we calculated the formation energies of defects in a variety of host materials:
$V_{\rm Zn}^{-2}$, $V_{\rm O}^{+2}$, and the Zn interstitial at the octahedral site (Zn$_i^{+2}$) in ZnO,~\cite{PhysRevB.77.245202,Oba:2011wz,Oba:2010gb,Lany:2008gk,PhysRevLett.93.225502}
$V_{\rm Mg}^{-2}$ and $V_{\rm O}^{+2}$ in MgO,~\cite{PhysRevLett.93.225502,PhysRevB.76.184103}
$V_{\rm Al}^{-3}$ and $V_{\rm O}^{+2}$ in Al$_2$O$_3$,~\cite{Choi:2013wi,PhysRevLett.93.225502} 
$V_{\rm Hf}^{-4}$ and $V_{\rm O}^{+2}$ on the three-fold coordinated O site in HfO$_2$,
$V_{\rm B}^{-3}$ and a defect complex of Ce on the N site coupling with neighboring four B vacancies (Ce$_{\rm N}$-4$V_{\rm B}^{-6}$) in c-BN,~\cite{PhysRevLett.110.065504}
Si$_i^{+2}$ and $V_{\rm Si}^{+2}$ in Si~\cite{Puska:1998hl,Corsetti:2011ix,Taylor:2011ue},
$V_{\rm As}^{+3}$ in GaAs,~\cite{Lany:2008gk,Lany:2009vf} and
$V_{\rm C}^{+2}$ in diamond~\cite{PhysRevB.71.035206}
that cover a wide range of crystal structures, local structures, chemistry (covalency and ionicity), and defect type (vacancies, interstitials, and substitutional impurities).
We checked these defects do not have delocalized perturbed host states with and without electron occupation for donorlike and acceptorlike states, respectively, 
which is a prerequisite of the electrostatics-based corrections including the FNV scheme;
perturbed host states require special treatments, e.g. by considering effective defect charges.~\cite{PhysRevB.77.245202,Komsa:2012ji}
The uncorrected and corrected defect formation energies with the PC model and extended FNV scheme are shown in Fig.~\ref{list_defect_formation_energy}.
The uncorrected defect formation energies are extrapolated to the dilute limit by fitting a function of the form $aN_{\rm atoms}^{-1}+bN_{\rm atoms}^{-1/3}+c$,
where $N_{\rm atoms}$ is the number of atoms in the supercell before introducing a defect.
We find that the cell size dependences of the FNV corrected defect formation energies with large supercells are extremely small, 
indicating the validity as the reference energies for measuring the errors.

\begin{figure*}
  \includegraphics[width=1\linewidth]{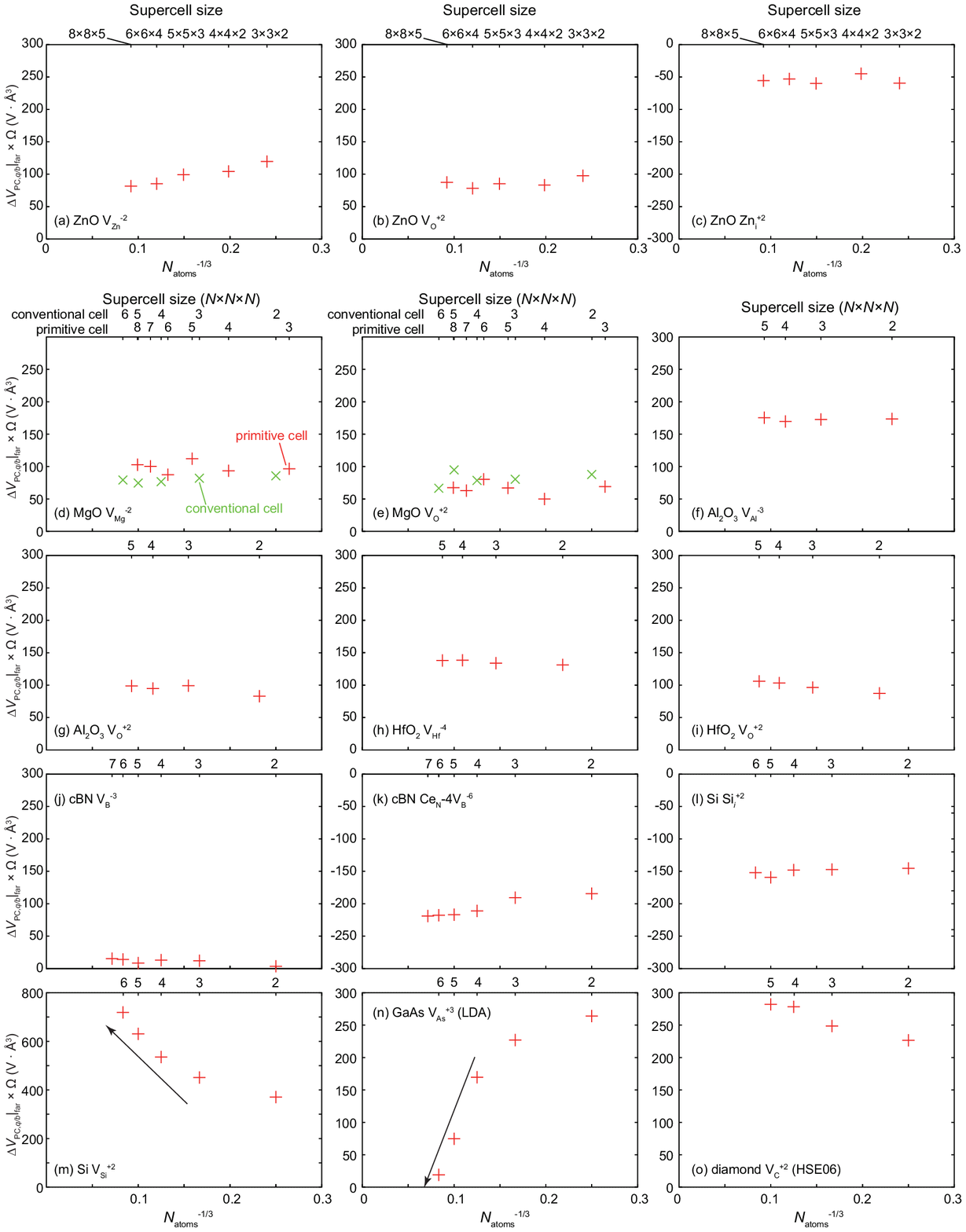}
  \caption{ $\Delta V_{{\rm PC},{q/b}}|_{\rm far} \cdot \Omega$ of the defects shown in Fig.~\ref{list_defect_formation_energy}.
    For comparison, scales of the horizontal axes are set to be the same except for $V_{\rm Si}^{+2}$ in Si.
    Large changes of $\Delta V_{{\rm PC},{q/b}}|_{\rm far} \cdot \Omega$ are indicated by arrows for guides to the eye.
 }
  \label{list_alignmentliketerm}
\end{figure*}

The PC correction basically improves the defect formation energies. 
Especially, $V_{\rm Zn}^{-2}$ and Zn$_i^{+2}$ in ZnO and $V_{\rm B}^{-3}$ in c-BN are well corrected.
However, it overshoots the energy of $V_{\rm O}^{+2}$ in ZnO, MgO, Al$_2$O$_3$, and HfO$_2$, $V_{\rm Si}^{+2}$ in Si, $V_{\rm As}^{+3}$ in GaAs, and $V_{\rm C}^{+2}$ in diamond.
The FNV correction, which is the sum of the PC correction and the alignmentlike term, greatly improves the defect formation energies in most cases, 
but $E[V_{\rm Zn}^{-2}]$ and $E[{\rm Zn}_i^{+2}]$ in ZnO are overshot.
$E[V_{\rm C}^{+2}]$ in diamond calculated with the HSE06 hybrid functional are also well corrected by the FNV scheme.

As discussed in Sec.~\ref{image}, an essential assumption is that the defect charge is encased in the supercell, 
and its distribution does not have cell size dependence.
The absence of the delocalized perturbed host states is just an essential condition and not sufficient to confirm this assumption, 
and the cell size dependence of $E[V_{\rm Si}^{+2}]$ may reflect violation of the assumption.
To check this, we plotted $\Delta V_{{\rm PC},{q/b}}|_{\rm far} \cdot \Omega$ in Fig.~\ref{list_alignmentliketerm}.
Supposing that the defect charge remains the same in different supercells,  $\Delta V_{{\rm PC},{q/b}}|_{\rm far} \cdot \Omega \approx \frac{2 \pi Q}{3\epsilon}$ 
must be constant because $Q$ is constant.

One can see that $\Delta V_{{\rm PC},{q/b}}|_{\rm far} \cdot \Omega$ are positive in vacancies and negative in interstitials.
This can be qualitatively understood as follows.
Supposing the unrelaxed geometry, Table~\ref{alignmentlike_table} shows the sign of second radial moment $Q$ and alignmentlike term $-q \Delta V_{{\rm PC},{q/b}}|_{\rm far}$ for charged vacancies and interstitials.
In case of vacancies, the valence and core electrons of the removed atom are also removed, and hence $\rho_d ({\bm r} \neq {\bm 0}) > 0$ and $Q=\int r^2 \rho_d ({\bm r}) d{\bm r} > 0$,
because the nucleus of the removed atom located at the defect site ${\bm r}={\bm 0}$.
On the contrary, in case of interstitials, due to the electrons of the interstitial atom, $\rho_d ({\bm r} \neq {\bm 0}) < 0$ and $Q<0$.
The alignmentlike term $-q \Delta V_{{\rm PC}, {q/b}}|_{\rm far}$ of negatively (positively) charged vacancies or positively (negatively) charged interstitials 
is then positive (negative) as listed in Table.~\ref{alignmentlike_table}.

Interestingly, $\Delta V_{{\rm PC},{q/b}}|_{\rm far} \cdot \Omega$ of $V_{\rm O}^{+2}$ are around 100 [$V \cdot \AA^3$] in any binary oxides in this study, 
and that of $V_{\rm B}^{+3}$ in c-BN is very small, reflecting a significantly small B$^{3+}$ ionic radius.
$\Delta V_{{\rm PC},{q/b}}|_{\rm far} \cdot \Omega$ are almost constant except for $V_{\rm Si}^{+2}$ in Si and $V_{\rm As}^{+3}$ in GaAs;
their $\Delta V_{{\rm PC},{q/b}}|_{\rm far} \cdot \Omega$ relate to the change of the defect charge distribution.
The behavior of $V_{\rm Si}^{+2}$ is notorious; its atomic and electronic structures and energetics strongly depend on the supercell size and $k$-points sampling.~\cite{Puska:1998hl,Corsetti:2011ix}
In fact, $\Gamma$-only $k$-point is not sufficient even with a 1726-atom supercell, and Monkhorst-Pack~\cite{PRB.5188} $2\times2\times2$ $k$-point mesh was adopted in this study.
This would be because the defect charge immersed in the valence band spread widely, and it leads to the erroneous defect-defect interactions.
Indeed, planar-averaged $\Delta V_{{\rm PC},{q/b}}|_{\rm far}$ in the unrelaxed geometry does not reach plateau between the defect and its image.
As a result, $\Delta V_{{\rm PC},{q/b}}|_{\rm far} \cdot \Omega$ increases as the supercell gets larger and larger, and more defect charge is encased.
On the other hand, $E[V_{\rm As}^{+3}]$ in GaAs is well corrected with the FNV scheme even with small supercells.
This may be because the defect states perturbed by the spurious potential are very similar in energy to the isolated defect state.
Then, the defect formation energy can be well calculated although the defect charge distribution does not converge.

For $V_{\rm Mg}^{-2}$ and $V_{\rm O}^{+2}$ in MgO, we calculated the defect formation energies with two types of supercells constructed from conventional and primitive cells, respectively,
which have simple cubic (sc) and face-centered-cubic (fcc) defect allocations.~\cite{Lany:2009vf}
Intuitively, the fcc supercells seem suited for defect calculations since the defect-defect distance is longer than that of the sc supercells in the same volume 
because of the larger coordination number in the fcc allocation.
Both $E[V_{\rm Mg}^{-2}]$ and $E[V_{\rm O}^{+2}]$ are, however, more accurately calculated with the sc supercells.
The reason is unclear but the defect-defect interactions might be enhanced in fcc supercells.
Such behavior has also been observed in $V_{\rm As}^{+3}$ in GaAs.~\cite{Lany:2009vf}

\begin{figure}
  \includegraphics[width=1\linewidth]{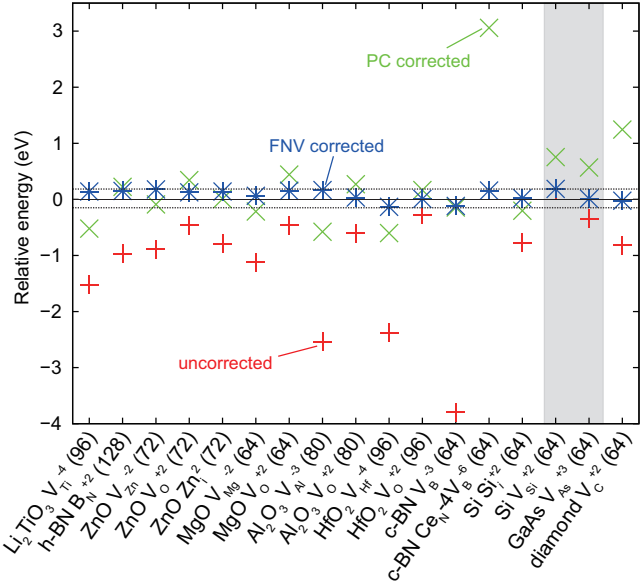}
  \caption{Relative defect formation energies estimated with supercells containing less than 100 atoms except for B$_{\rm N}^{+2}$ antisite defects in h-BN,
           compared to FNV corrected energies estimated with the largest supercells.
           Atomic relaxation is considered in any cases.
           The uncorrected energy of Ce$_{\rm N}$-4$V_{\rm B}^{-6}$ in c-BN is -11.6 eV.
           The defects in the shaded area have large cell size dependences of $\Delta V_{{\rm PC},{q/b}}|_{\rm far} \cdot \Omega$,
           indicating violation of the assumption that the defect charge distribution in the supercell is the same as the isolated one.
           Note that the corrected energies with the extended FNV scheme are located within $\pm 0.19$ eV. }
  \label{absolute}
\end{figure}

The absolute error is of importance in practice, and thus we plot the relative defect formation energies calculated 
with small supercells containing around 100 atoms in Fig.~\ref{absolute}.
Such small supercells are convenient for computationally expensive calculations.
It is found that the defect formation energies are excellently corrected by the extended FNV scheme and the differences from those in the dilute limit are less than 0.19 eV in our test set.
Surprisingly the errors do not largely depend on the defect charge as Freysoldt {\it et al.} pointed out in Ref.~\onlinecite{Freysoldt:2009ih}.

\begin{figure*}
  \includegraphics[width=1\linewidth]{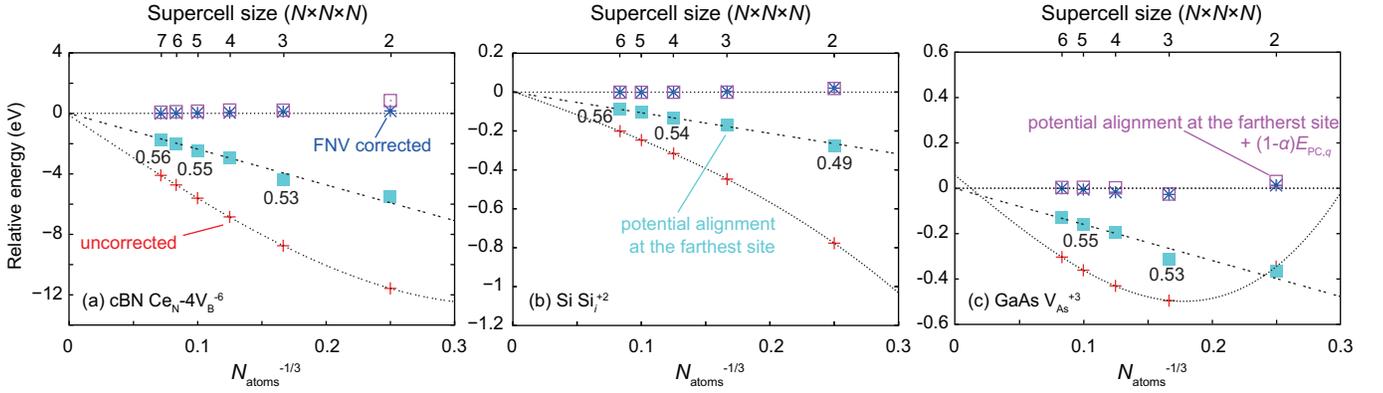}
  \caption{Relative formation energies of (a) Ce$_{\rm N}$-4$V_{\rm B}^{-6}$ in c-BN, (b) Si$_i^{+2}$ in Si, (c) $V_{\rm As}^{+3}$ in GaAs,
    and their energies corrected with the FNV scheme, conventional potential alignment, and potential alignment plus $(1-\alpha({\bm r})) E_{\rm PC}$.
    The potential alignment was performed at the farthest atomic site.
    In cases where the farthest atoms locate at (0.5 0.5 0.5), $\alpha=0.57$.
    Otherwise, $\alpha$ values are shown.
 }
  \label{conventional_alignment}
\end{figure*}

The conventional potential alignment discussed in Sec.~\ref{alignment_revisited} is also reviewed 
with Ce$_{\rm N}$-4$V_{\rm B}^{-6}$ in c-BN, Si$_i^{+2}$ in Si, and $V_{\rm As}^{+3}$ in GaAs in Fig.~\ref{conventional_alignment}.
The energies corrected with the potential alignment at the farthest atomic site from the defect and its images have nearly linear dependence against $L^{-1}$.
The deviations from the linear dependence are, however, larger than that of the Si$^{+}$ ion shown in Fig.~\ref{point_charge_model}(b).
This is because the farthest atoms do not always locate at (0.5 0.5 0.5) of the supercells.
For instance, such atoms locates at (0.5 0.5 0.5) when $N$ in $N^3$-fold Si$_i$ supercell is odd number, but it does not when $N$ is even number.
One can find that a sum of the conventional potential alignment and $1 - \alpha$ of the 
PC correction energy almost recovers the FNV corrected energies, consistent with the Si ionization energy in Sec~\ref{alignment_revisited}.
The remaining differences observed in Fig.~\ref{conventional_alignment} correspond to the difference in potential sampling methods;
the potential alignment is performed at the farthest atomic site whereas the FNV correction is performed with the potential in the sampling region in this study.

The FNV scheme can correct the defect formation energies up to the $L^{-3}$ order.
We here discuss the origins of the remaining error.
The error sources considered are as follows:
(i) The defect charge spills out from the supercell. 
(ii) The defect charge distribution is affected by the spurious potential caused by the defect charges and background charge.~\cite{Makov:1995vk}
(iii) Sampling error for the potential alignmentlike term.
(iv) Correction energy with $L^{-5}$ or higher orders.
(v) Defect-induced dipoles, which contributes to decrease the formation energy as shown in Eq.~(\ref{Makov-Payne}).
(vi) Defect-induced elastics, which contributes to increase the formation energy.

(i) -- (iii) can be checked with $\Delta V_{{\rm PC},{q/b}}|_{\rm far} \cdot \Omega$ as shown in Fig.~\ref{list_alignmentliketerm}.
(iv) would be dominant when the fitting with a function of the form $aN_{\rm atoms}^{-1}+bN_{\rm atoms}^{-1/3}+c$ works poorly.
(v) may be only related to the defects in ZnO without an inversion symmetry in the present study.
The omission of the leading dipole term, however, should underestimate the defect formation energies and this is not true for the defects in ZnO.
Roughly estimating the dipole energy, we calculate the dipole energy of the charges $\pm e$ separated $0.5~\AA$
from each other (${\bm p} = 0.5 \cdot {\bm z}$ [$e \cdot \AA$] ) in the $4 \times 4 \times 2$ supercell of ZnO ($\Omega= 1590~\AA^3$) 
with a theoretical dielectric constant $\epsilon = \langle \epsilon_{ii} \rangle =10.64$, and obtain $\frac{2 \pi {\bm p}^2}{3\epsilon \Omega} = 5.6$ meV,
which is negligibly small compared to the remaining error of the defects in ZnO.
Only (vi) is not explicitly dependent on $q$.
Since the remaining errors after the extended FNV correction are not strongly dependent on the defect charges, (vi) might be a main error source for the defects localized in the supercell.
Note that the lattice optimization of the defective supercell is not useful to reduce the elastic energy in general,
because the spurious elastic interactions occur under periodic boundary conditions, and can underestimate the defect formation energy.
A combination of first-principles calculations and elastic theory might resolve this issue.~\cite{PhysRevB.88.134102}

\begin{table}[t]
 \caption{Sign of charge $q$, second radial moment $Q$, and alignmentlike term $-q \Delta V_{{\rm PC},{q/b}}|_{\rm far}$ for charged vacancies and interstitials in a A$^{+N}$B$^{-N}$ binary compound.}
  \begin{center}
  \begin{threeparttable}
  \begin{tabular}{ccccc}\\ 
  \hline
  \hline
     & $V_{\rm A}^{-N}$ & $V_{\rm B}^{+N}$ & $X_i^{+N}$ & $X_i^{-N}$ \\
  \hline
  {\it q} & - & + & + & - \\
  {\it Q}, $\Delta V_{{\rm PC},{q/b}}|_{\rm far}$ & + & + & - & - \\
  $-q \Delta V_{{\rm PC},{q/b}}|_{\rm far}$ & + & - & + & - \\
  \hline
  \hline
  \end{tabular}
 \end{threeparttable}
 \end{center}
 \label{alignmentlike_table}
\end{table}

\section{Conclusions}\label{conclusion}
In this paper, we have discussed the electrostatics-based finite cell size corrections for first-principles point defect calculations under periodic boundary conditions.
In the beginning, the PC correction that is the leading term of the image-charge correction has been reviewed in detail.
Then, we have introduced the higher order correction term O($L^{-3}$) derived by the MP and FNV schemes.
We then have proposed a way to extend the FNV scheme to be applicable to a wide variety of materials. 
Firstly, we have introduced atomic site potential for determining the potential offset between the defect-induced potential and PC potential, and compared it with the planar-averaged potential.
Secondly, we have introduced a PC model with the anisotropic form for evaluating long-range Coulomb interactions.
The FNV scheme with the anisotropic form has been tested with $V_{\rm Ti}^{-4}$ in $\beta$-Li$_2$TiO$_3$ and B$_{\rm N}^{+2}$ in h-BN, 
and it is found that their formation energies are well corrected by the extended FNV scheme.

The potential alignment, which has been discussed by many authors for long time, has also been revisited in Sec.~\ref{alignment_revisited}.
We have concluded that the potential alignment is unnecessary when the image-charge correction is properly considered.
This is confirmed by calculating the ionization energy of the Si atom.
We also have discussed the physical meaning of the conventional potential alignment, 
and found that it contains a part of the PC correction energy and full of the potential alignmentlike term of the FNV scheme.
We propose that this would be the origin of the absence of the $L^{-3}$ order term after applying the potential alignment previously reported.~\cite{Lany:2009vf}
The amount of the PC correction energy included by the potential alignment depends on the coordinates where the potential alignment is attained.

In Sec.~\ref{test}, we have tested the accuracy of the extended FNV scheme with a test set composed of 17 defects in 10 materials, 
and found that it systematically improves the defect formation energies.
The signs of the second radial moment $Q$ and alignmentlike term $-q\Delta V_{{\rm PC}, {q/b}}|_{\rm far}$ have also been discussed.
The corrected defect formation energies with -6 to +3 charges calculated with around 100-atom supercells are within $\pm$ 0.19 eV compared to those in the dilute limit.
We believe that the extended FNV scheme is a powerful tool for correcting defect formation energies as long as the defect charges are encased in the supercells.

\begin{acknowledgments}
We thank Atsuto Seko and Minseok Choi for valuable discussions.
This work was supported by the MEXT Elements Strategy Initiative to Form Core Research Center
“Tokodai Institute for Element Strategy (TIES)” and a Grant-in-Aid for Scientific Research on Innovative Areas "Nano Informatics" (grant number 25106005) from JSPS.
Computing resources of ACCMS at Kyoto University were used in this work.
The visualization of crystal structures were performed with VESTA.~\cite{JApplCryst.41.653}
\end{acknowledgments}


\end{document}